\documentclass[doublecol]{epl2}
\usepackage{amsmath} 

\title{Vacuum fluctuation forces between ultra-thin films}

\author{A. Benassi\inst{1,2} \and C. Calandra\inst{2}}
\shortauthor{A. Benassi \and C. Calandra}

\institute{                    
  \inst{1} 
CNR/INFM-National Research Center on nanoStructures and bioSystems at Surfaces (S3)\\
Via Campi 213/A, I-41100 Modena, Italy \\
  \inst{2} 
Dipartimento di Fisica, Universit\'a di Modena e Reggio Emilia\\
Via Campi 213/A, I-41100 Modena, Italy}

\pacs{12.20.Ds}{QED specific calculations}
\pacs{68.65.Fg}{Quantum wells}
\pacs{64.60.an}{Finite-size systems}

\abstract{We have investigated the role of the quantum size effects in the evaluation of the force caused by electromagnetic vacuum fluctuations 
between ultra-thin films, using the dielectric tensor derived from the particle in a box model. Comparison with the results obtained by adopting a 
continuum dielectric model shows that, for film thicknesses of $1\div 10$ nm, the electron confinement causes changes in the force intensity with 
respect to the isotropic plasma model which range from  $40$\%  to few percent depending upon the film electron density and the film separation. The 
calculated force shows quantum size oscillations, which can be significant for film separation distances of several nanometers. The role of
electron confinement in reducing the large distance Casimir force is discussed.}
 
\begin{document}
\maketitle
\section{Introduction}
Quantum mechanical forces, namely van der Waals and Casimir forces, between nano objects are expected to be important in the operation of micro and 
nano systems \cite{serry,buks,zhao}. These forces depend significantly upon the optical properties of the interacting 
objects: for interacting films, variations in the experimental optical parameters, caused by film morphology, can determine a change in the force of 
the order of $10$\% \cite{pirozhenko,iannuzzi}. Since the basic work from Lifshitz and coworkers \cite{lifshitz,dzyaloshinskii} theoretical studies have been focused mainly on the determination 
of the forces between slabs, including semi-infinite slabs, on the basis of a continuum description of the material dielectric function 
\cite{Bordag,bentsen,lambrecht2,geyer}. This is a 
simplified description of a film, that neglects the modifications in the electronic structure related to the boundaries. It is expected to hold when 
the size of the film is large, so that the surfaces play a minor role in determining the dielectric response. For metallic films the 
calculated electronic distributions deviate significantly from the bulk behaviour when the size of the film is less than ten times the 
Fermi wavelength \cite{rogers1,rogers2}. If the size of the film is of the order of few nanometers the continuum model does not 
provide an accurate description of the film properties and boundary effects cannot be neglected. Such effects arise as a consequence of the discretization of the energy bands due to the confinement potential, which 
produces the quantization of the electron energy levels in various sub-bands and affects the optical and electrical properties \cite{loly,lindgren1,lindgren2,chiang,jalochowski1,jalochowski2,rogacheva,hens}. 
\section{The quantum model}
The characteristics features induced by the quantization can be described by a model in which independent electrons of mass $m$ are confined within a distance $d$ in the direction of the surface normal (particle in a box model) \cite{wood,czoschke}. Assuming a jellium model and perfect planar surfaces the eigenvalue spectrum is simply given by:
\begin{equation}
E_{\textbf{k}_{\parallel},n}=\frac{\hbar^2}{2 m}(k^2_{\parallel}+k^2_{\perp})=\frac{\hbar^2}{2 m}(k^2_{\parallel}+\frac{n^2\pi^2}{d^2})
\end{equation}
i.e. described by a continuous quantum number $k_{\parallel}$, giving the modulus of the parallel wavevector, and a discrete sub-band index $n$ coming from the quantization of the perpendicular wavevector $k_{\perp}$. The corresponding wavefunctions are:
\begin{equation}
\psi_{\textbf{k}_{\parallel},n}(\textbf{r})=\sqrt{\frac{2}{V}}e^{i\textbf{k}_{\parallel}\cdot\textbf{r}_{\parallel}}sin\bigg(\frac{n \pi}{d}z\bigg)
\label{wfc}
\end{equation}
where $V$ is the volume of the quantum well given by the product of the well surface $A$ and the well thickness $d$.
In this simple model the electrons behave as a two-dimensional gas along the $x$ and $y$ space directions and as standing waves in the $z$ direction, 
with nodes on the boundaries. As a first approximation one can assume the size $d$ of the quantum well to be the same as the size $D$ of the ion 
distribution of the film. This approximation is too crude since it does not allow the spilling of the electron density past the film boundaries given by the positive charge distribution, thus leading to a depletion of negative charge near the surfaces. One way to eliminate this 
inconvenience without introducing softer boundaries is to allow the electron charge to be distributed on a larger size than the positive charge. It 
is common practice to discuss the effects of the confining potential in deposited films using the phase accumulation model \cite{echenique}, according 
to which the condition that is satisfied by a quantum well state is
\begin{equation}
2 k_{\perp} d +\phi_{A}+\phi_{B}=2\pi n
\end{equation}
where $\phi_A$ and $\phi_B$ are the phases of the eigenfunctions accumulated at the two film interfaces and $n$ are 
integer numbers. For a free standing film, limited by two vacuum-film interfaces, $\phi_A=\phi_B$. 
The case of an infinite quantum well is recovered by 
simply imposing $\phi_A=-\pi$. For softer profiles of the confinement potential the phase shifts are expected to increase and to depend upon the energy. 
One can introduce a more realistic description of the film electron states, still using the infinite potential well, by allowing the effective 
width $d$ to be larger than the size of the ion distribution $d= D + 2\Delta$, thus obtaining the quantization condition $D + 2 \Delta_n = n\pi/k_{\perp}$, where $\Delta_n$ gives the 
shift in the potential well width that allows to reproduce the charge spilling out of the $n$-th state. This introduces an energy dependence of the 
shift that in principle could be obtained by fitting the quantum well energies to experimental data or to the energy level distribution resulting 
from a first principle calculations. 
Previous works \cite{otero1,otero2} have shown that the energy spectrum is very sensitive to the position of the barriers and relatively insensitive to the barriers 
height.
For the present preliminary study we take an averaged $d$ value obtained by simply imposing the film to be neutral with an electron charge of size 
$d$ and a positive charge of size $D$. This leads to the following expression for the effective film thickness:
\begin{equation}
d=\frac{D}{G(m_{F})}
\label{dd}
\end{equation}
with: 
\begin{equation}
G(m_F)= \frac{3m_0}{2m_F}\bigg[1-\frac{( m_0+1)(2 m_0+1)}{6 m_{F}^{2}}\bigg]
\end{equation}
where $m_F=k_Fd/\pi$, $k_F$ is the Fermi wavevector that is related to the ion electron density $N_0$ by the relation $k_{F}^{3}= 3\pi^2N_0$, and $m_0$ is the integer part of $m_F$. Notice that, once $D$ and the positive charge density $N_0$ are given, the $d$ value can be determined unambiguously\footnote{For the calculation of the film charge density see for example reference \cite{czoschke}.}. For large $D$, when $m_0 \simeq m_F$ , one obtains $d = D + 3 \pi/4 k_F$, which shows that the spilling out of the charge in this limit is proportional to the Fermi wavelength.\\ 
With respect to the classical description of thin film optical and transport properties \cite{reute,hutchinson}, where surface effects are expressed as boundary conditions on the electron distribution function in terms of the mean free path and the fraction of the electrons scattered specularly at the surface, this theory has the surface effects incorporated as boundary conditions in the one-electron hamiltonian, the only parameters being the film size and the Fermi energy. 
\section{The model dielectric tensor}
\begin{figure}
\onefigure[width=8.5cm]{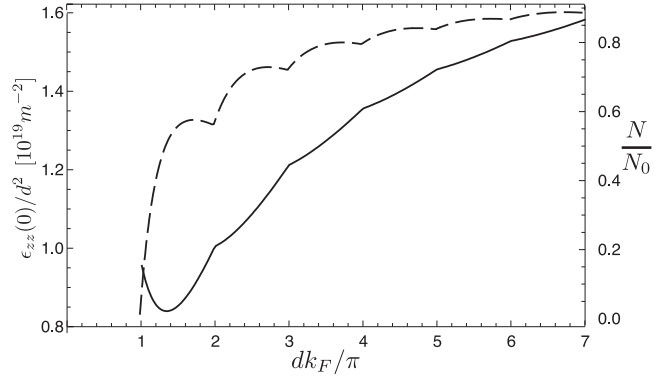}
\caption{Static value of the $zz$ component of the dielectric tensor (continuous line) and ratio between electron densities with and without quantum confinement (dashed line) as a function of $d\: k_F/ \pi$ for a film with $\Omega_p=10^{16}$ rad/sec. In both cases kinks occur, with discontinuities in the derivative, at $k_F/\pi$ multiples.}
\label{fig1}
\end{figure}
The derivation of the dielectric function of the film described by the previous model, leads to a description of the optical properties
that accounts for sub-bands transitions \cite{wood}. The finite extension of the film along the $z$ direction implies the anisotropy of its dielectric response with the diagonal components:
\begin{equation}
\begin{split}
\epsilon_{\alpha\alpha}(\omega)&=1-\frac{\omega_{p}^{2}}{\omega^{2}}-\frac{4\pi e^{2}}{A d m^{2}
\omega^{2}}\sum_{{\bf k}_{\parallel},n}\sum_{{\bf k}_{\parallel}',n'} \times \\
&\times\frac{f(E_{\bf{k}_{\parallel},n})-f(E_{\bf{k}_{\parallel}',n'})}
{E_{\bf{k}_{\parallel},n}-E_{\bf{k}_{\parallel}',n'}-\hbar\omega}\vert\langle\psi_{{\bf k}_{\parallel},n}\vert\hat{p_{\alpha}}\vert\psi_{{\bf k}_{\parallel}',n'}
\rangle\vert^{2}
\end{split}
\label{dieltens}
\end{equation}
here $\alpha=x,y,z$ labels the cartesian component of the tensor, $\hat{p}_{\alpha}$ indicates the component of the electron linear momentum,
$\omega_{p}=\Omega_{p} N/N_0$ is the plasma frequency of the quantized electron gas ($\Omega_{p}=\sqrt{4\pi e^2 N_0/m}$ is the free electron plasma frequency) and $f(E_{\bf{k}_{\parallel},n})$ is the occupation factor of the $(\bf{k}_{\parallel},n)$ state. The off-diagonal component are equal to zero. This expression differs from the 
plasma model dielectric function adopted in previous studies in that: 
(i) it has a tensor character with $\epsilon_{xx}=\epsilon_{yy}\neq\epsilon_{zz}$, (ii) the plasma frequency $\omega_{p}$ depends 
upon the film density $N$, which changes as a function of the film thickness, (iii) through the double sum in the second member it accounts for transitions between lateral sub-bands, whose probability amplitude is expressed by the momentum matrix element between the one electron wavefunctions (\ref{wfc}). It can be easily shown that these 
transitions do not affect the lateral components of the dielectric tensor, which are given by the simple expression of the plasma dielectric function
\begin{equation}
\epsilon_{xx}(\omega)=\epsilon_{yy}(\omega)=1-\frac{\omega_p^2}{\omega^2}
\end{equation}
because the momentum matrix element for $x$ and $y$ component vanishes.
They modify $\epsilon_{zz}(\omega)$ whose low frequency behaviour gives a finite dielectric 
constant: 
\begin{equation}
\epsilon_{zz}(0)=1+\frac{d m_F^2}{6 \pi^2 a_0}\bigg[15 \bigg(S_4-\frac{S_2}{m_F^2}\bigg)+\pi^2 \frac{(1-m_F S_2)}{m_F}\bigg]
\end{equation}
here $a_0$ is the Bohr radius, and :
\begin{equation}
S_k=\sum_{n=1}^{m_0}\frac{1}{n^{k}}
\end{equation}
The metallic character, with vanishing minimum excitation energy and divergent 
$\epsilon_{zz}(\omega)$ at low frequency, is recovered for $d$ going to infinity. The plot of  $\epsilon_{zz}(0)$, given in fig.\ref{fig1}, shows an 
oscillatory behaviour as a function of $d$ with periodicity given by $\pi/k_F$ superimposed over a regularly increasing curve. These oscillations 
arise from the periodic crossing of the Fermi energy by the sub-bands with increasing thickness. At each crossing a new sub-bands gets filled and 
optical transitions having such sub-band as initial state become possible. The occurrence of quantum size oscillations has been pointed out by several 
authors specially with reference to the electron density, the total electronic energy and the film electrical conductivity \cite{czoschke,czoschke2,wang}. Fig \ref{fig1} shows that they are distinct features of the dielectric response too.
\section{The force}
\begin{figure}
\onefigure[width=6.5cm]{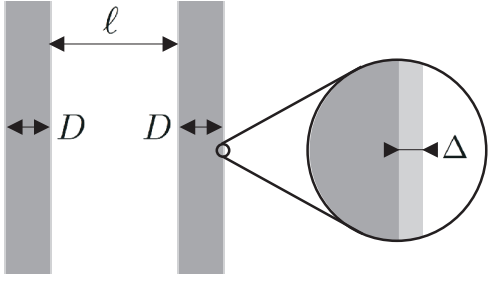}
\caption{Two identical interacting thin films, definition of $\ell$, $D$ and the electronic spill-out $\Delta$.}
\label{fig2}
\end{figure}
\begin{figure}
\onefigure[width=8.5cm]{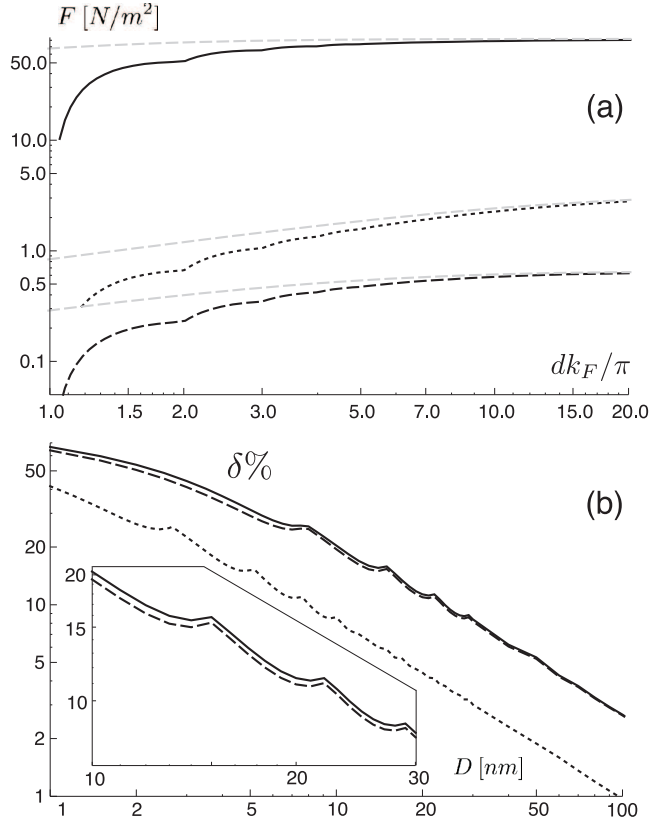}
\caption{(a) Black lines represent $F_{Q}$ as a function of the film thickness for different film densities and separations, $\Omega_{p}=10^{14}$ 
rad/sec and $\ell=10$ nm (continuous line), $\Omega_{p}=5\cdot 10^{14}$ rad/sec and $\ell=50$ nm (dotted line), $\Omega_{p}=10^{14}$ rad/sec and 
$\ell=50$ nm (dashed line). The gray dashed lines represent $F_{P}$ for the same parameters. (b) relative percentual difference between the force with and without quantum size effects.}
\label{fig3}
\end{figure}

The expression of the force per unit area $F$ at $T=0^{\circ}$ K in the configuration illustrated in the inset of fig.\ref{fig2} can be obtained by extending previous results relative to isotropic films \cite{Bordag,zhou} to the case of films with anisotropic dielectric tensor. One obtains:
\begin{gather}
\nonumber
F=-\frac{\hbar }{2 \pi^2}\int_{0}^{\infty} kdk\int_{0}^{\infty}d\omega \gamma (\omega)
\bigg[\frac{Q_{TM}(i\omega)^2}{1-Q_{TM}(i\omega)^2}+\\
+\frac{Q_{TE}(i\omega)^2}{1-Q_{TE}(i\omega)^2}\bigg] 
\end{gather}
\begin{gather}
Q_{TM}=\frac{\rho_{TM}(1-e^{-2 \gamma_{TM} D})}{1-\rho_{TM}^2e^{-2 \gamma_{TM} D}}e^{-\gamma \ell}\\
Q_{TE}=\frac{\rho_{TE}(1-e^{-2 \gamma_{TE} D})}{1-\rho_{TE}^2e^{-2 \gamma_{TE} D}}e^{-\gamma \ell}\\
\rho_{TM}=\frac{\gamma_{TM}(\omega)-\gamma\epsilon_{xx}(\omega)}{\gamma_{TM}(\omega)+\gamma\epsilon_{xx}(\omega)}    
\:\:\:\:\rho_{TE}=\frac{\gamma_{TE}(\omega)-\gamma(\omega)}{\gamma_{TE}(\omega)+\gamma(\omega)}
\end{gather}
\begin{gather}
\nonumber
\gamma(\omega)=\sqrt{k^2-\frac{\omega^2}{c^2}}\qquad \gamma_{TE}(\omega)=\sqrt{k^2-\frac{\omega^2}{c^2}\epsilon_{xx}(\omega)}\\
\gamma_{TM}(\omega)=\sqrt{\bigg(\frac{k^2}{\epsilon_{zz}(\omega)}-\frac{\omega^2}{c^2}\bigg)\epsilon_{xx}(\omega)}
\end{gather}
This equation, which to the authors khowledge has never been used previously for vacuum fluctuation force calculations, deserves a few 
comments. (i) it has been obtained by considering the zero point energies associated with the electromagnetic modes of films of finite thickness 
under the assumption that the dielectric permittivity is represented by an anisotropic diagonal tensor. It differs from the original Lifshitz formula 
both because it depends upon the film size $D$ (i.e. has been obtained with the electromagnetic field boundary conditions appropriate to a finite 
size film and not for a semi-infinite system) and for the presence of the anisotropic permittivity. (ii) the force goes to zero as the film size $D$ 
vanishes, since both $Q_{TM}$ and $Q_{TE}$ go to zero. (iii) to calculate the force one has to determine the frequency dependence of the dielectric 
tensor, which we have obtained from equation (\ref{dieltens}) using the eigevalues and the wavefunctions of the quantized film. This approach differs 
from what is 
commonly done in dispersion forces calculations based on a continuum description of the dielectric properties, where the dielectric function is 
derived from empirical expressions which do not bear a direct relation to the electronic band structures of the interacting bodies \cite{Bordag,bentsen,lambrecht2,geyer}.
In the following the force per unit area calculated for the dielectric tensor (\ref{dieltens}) appropriate to the film is denoted by $F_Q$, while 
$F_P$ indicates the force calculated using the isotropic plasma model i.e. $\epsilon_{xx}=\epsilon_{yy}=\epsilon_{zz}$. To evaluate the importance of 
the size quantizations, we are interested to compare $F_Q$, calculated for given film thickness $D$ and ion density $N_0$ at different distances 
$\ell$, 
with $F_P$ calculated in the same configuration. 
Fig.\ref{fig3} (a) displays curves of $F_Q$ calculated at fixed $\Omega_p$ and $\ell$ values as a function of the film thickness. The $\Omega_p$ 
values 
correspond to free carrier densities of heavily doped semiconductors. These systems are those which display higher modifications.
We have considered distances $\ell$ ranging from 
$10$ to $50$ nm. For comparison the curves of  the corresponding $F_P$ vales are reported. One can see significant differences between the two models 
for $d$ values that are of the order of few multiple integer of the half of the Fermi wavelength. The curves show quantum size oscillations with the 
expected periodicity $k_Fd=n\pi$ superimposed over a regularly increasing behaviour. For thick films the results of the isotropic plasma model are 
recovered. 
The size induced modifications are better illustrated by the quantity: 
\begin{equation}
\delta=\frac{F_P-F_Q}{F_P}
\label{percdiff}
\end{equation}
which gives the relative variation of the force with respect to the isotropic plasma model. Fig \ref{fig3} (b) presents plots of $\delta$ for the 
cases under consideration. It is seen that the size induced modifications are very large, ranging from $50\%$ to $10\%$, for films of nanometric 
thickness even at distances $\ell$ of several nanometers.\\
\begin{figure}
\onefigure[width=8.5cm]{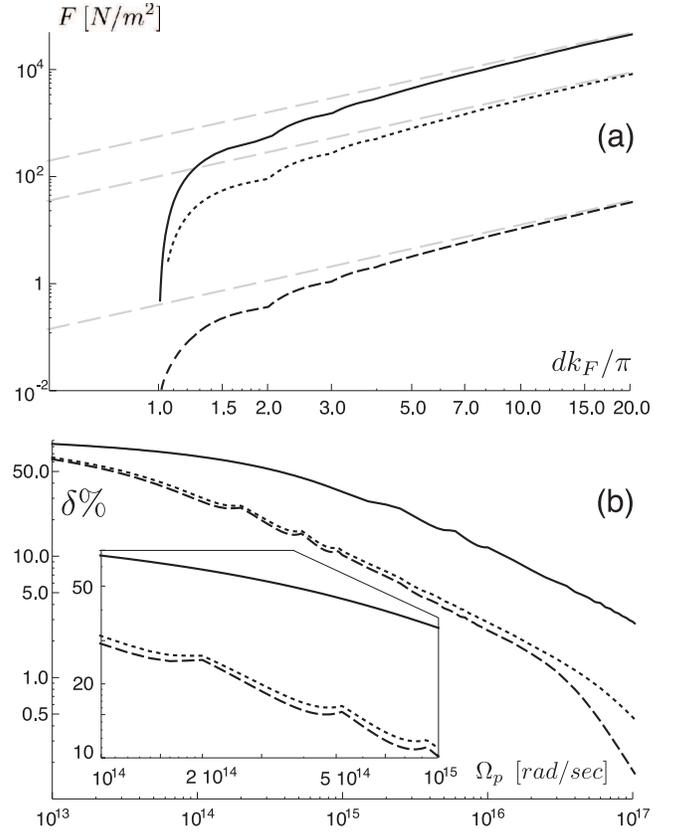}
\caption{(a) Black lines represent $F_{Q}$ as a function of the plasma frequency for different film thicknesses and separations, $D=1$ nm 
and $\ell=10$ nm (continuous line), $D=5$ nm and $\ell=10$ nm (dotted line), $D=5$ nm and $\ell=50$ nm (dashed line). The gray 
dashed lines represent $F_{P}$ for the same parameters. (b) relative percentual difference between the force 
with and without quantum size effects.}
\label{fig4}
\end{figure}
One can see similar modifications in the theoretical results obtained by keeping the film size constant and changing its density. This is illustrated 
in figs. \ref{fig4} (a) and (b) which display force curves obtained as a function of plasma frequency.
Notice that even for typical metallic densities ($\Omega_p$ of the order of $10^{15}\div 10^{16}$ rad/sec) the deviations from the plasma model are quite significant and can be of the order of several percents for film separation distance $\ell$ ranging from $10$ to $50$ nm.
\begin{figure}
\onefigure[width=8.5cm]{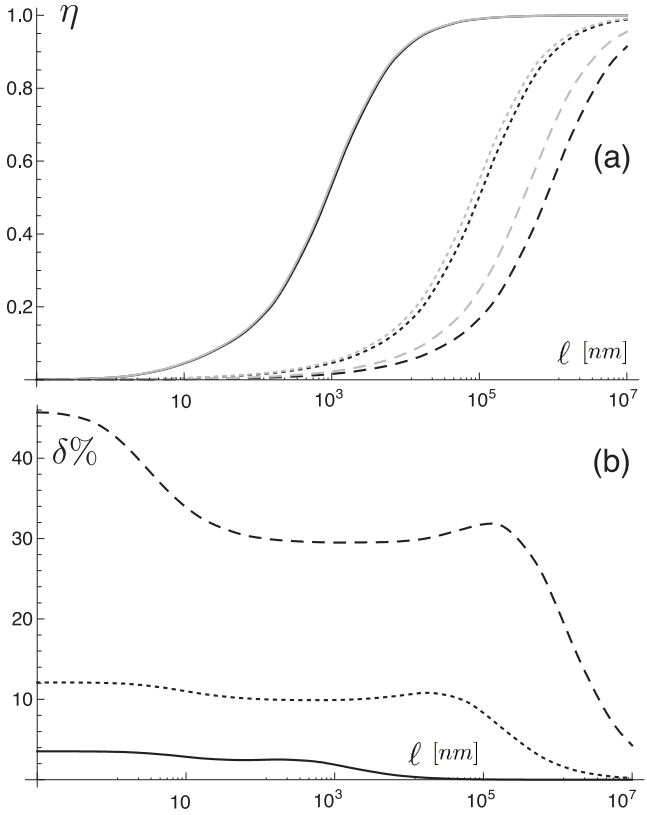}
\caption{(a) Black lines represent $\eta_Q$ as a function of the films distance $\ell$ for different densities and thicknesses: continuous line $\Omega_{p}=10^{16}$ rad/sec $D=5$ nm, 
 dashed line $\Omega_{p}=10^{15}$ rad/sec $D=1$ nm, dotted line $\Omega_{p}=10^{15}$ rad/sec $D=5$ nm.
Gray lines represent $\eta_{P}$ as a function of $\ell$ for the same parameters. (b) Percentual difference (\ref{percdiff}) for the three cases of the (a) plot.}
\label{fig5}
\end{figure}
In a previous study on the Casimir effect for metal and semiconductor slabs it has been pointed out that the Casimir force can be considerably reduced by decreasing the slab thickness \cite{lambrecht2}. It is interesting to see whether quantum confinement effects are important in determining the reduction at large $\ell$ values. To this end we plot in fig. \ref{fig5} (a) the reduction factors $\eta_{P} = F_{P}/F_{CAS}$ and $\eta_Q  = F_Q/F_{CAS}$ \cite{Lambrecht}, where 
\begin{equation}
F_{CAS}=-\frac{\hbar c \pi^2}{240 \ell^4}
\end{equation}                                                           
is the force between perfectly reflecting mirrors at separation distance $\ell$ \cite{casimir2}. We consider films of nominal thickness $D=1\div 5$ 
nm and different electron densities. One can see that the effects of the quantum confinement tend to decrease the reduction factor over a large 
distance range. The correction depends upon both the film size and the electron density, being larger for smaller values of these quantities.
Fig. \ref{fig5} (b) plots the calculated relative difference $\delta$ as a function of $\ell$ for the same cases. It shows that relative variations 
of the force remain significant even at very large distances thus giving sizeable reduction to the force in the large $\ell$ Casimir regime. Notice 
that $\delta$ does not show a monotonous behaviour as a function of $\ell$: it decreases regularly at short distances, it remains constant in a large 
range of $\ell$ values and, before going to zero for very large distances, it shows a local maximum that shifts at higher $\ell$ values upon decreasing the thickness 
and/or the electron density. This behaviour is quite typical and can be reproduced in several cases.
\begin{figure}
\onefigure[width=8.5cm]{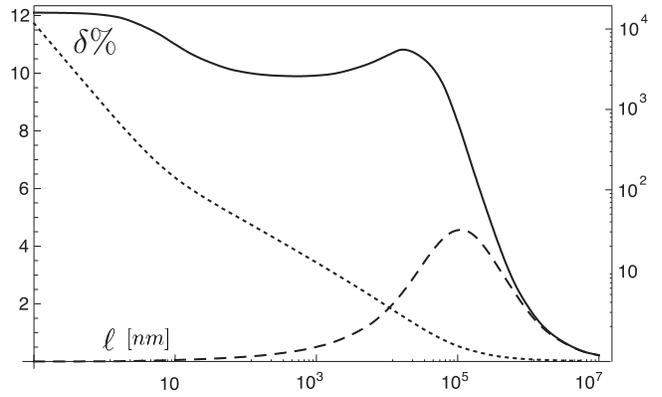}
\caption{Continuous line represents the percentual relative difference $\delta$, dashed line represents $\eta_P - \eta_Q$, dotted line represents the ratio $1/\eta_P$ for the case $\Omega_{p}=10^{15}$ rad/sec $D=5$ nm}
\label{fig6}
\end{figure}
To understand this behaviour we notice that $\delta$ can be written as 
\begin{equation}
\delta = 1 - \frac{\eta_Q}{\eta_P}
\end{equation}
Fig. \ref{fig6} shows the typical behaviour of the numerator and of $1/\eta_P$ as a function of $\ell$. The first curve decreases more or less 
regularly upon increasing the film separation. The change in the slope arises from different inverse power behaviour of the short range (dispersion 
force) compared to long range Casimir force \cite{lifshitz,dzyaloshinskii,Bordag,zhou}. The second curve shows a regular increase at short distances, it has a maximum and 
goes to zero at large $\ell$ values. Therefore the maximum in the $\delta$ corresponds to the range of $\ell$ values where the reduction of the 
Casimir force is larger compared to the ideal case, while the plateau is due to the combined effect of the decrease of $\eta_P - \eta_Q$ and the rise of 
the $1/\eta_P $ curve.
\section{Conclusions}
Using the particle in a box model we have presented results that show under what conditions the electron confinement can affect the quantum mechanical force between free-standing films and we have given evidence of the presence of quantum size oscillations effects for small thickness and/or low density films. In extending this theory to the more realistic case of deposited films, one has to account for the penetration of the electron wavefunctions into the substrate. This requires a modification of the quantization condition compared to the abrupt infinite barrier potential adopted in the present paper. However for semiconductor substrates where the confinement is essentially due to the fundamental gap, the modifications of the potential should not alter appreciably  our main conclusions. For metallic substrate the extension is less obvious since the confinement is related to the existence of symmetry or relative gaps of the projected bulk band structure, that exist for particular directions only. In this case the solid-film interface has to be simulated with a softer confinement potential \cite{ogando}.\\
In view of the sensitivity of our results to the film density, the observation of the effects we have reported could be made possible by some experimental technique that allows to modify the carrier density in overlayers (doping \cite{chen}, creating electron-hole plasma by illumination \cite{klimchitskaya2} etc.). Also by changing the substrate one may affect the confinement potential, making it more abrupt or more rounded, and enhance or reduce quantum size effects.\\
Beside including wavefunction penetration into the substrate, theoretical improvements to the present theory could come from the inclusion of (i) local field effects in the description of the film dielectric response \cite{li}, (ii) relaxation time into the metal dielectric function following the prescription appropriate to finite size systems \cite{garik}. Work along these lines is in progress.
\acknowledgments
AB thanks \emph{CINECA Consorzio Interuniversitario} ({\tt
www.cineca.it}) for funding his Ph.D. fellowship.
\bibliographystyle{unsrt}

\end{document}